\documentclass[12pt]{article}

\author{Vladimir A. Petrov\footnote{e-mail: Vladimir.Petrov@ihep.ru}}

\title{  Why the Bethe-West-Yennie Formula for Coulomb-Nuclear Interference Is Inconsistent. }
\date{}

\begin{document}

\maketitle
\begin{center}
A. A. Logunov Institute for High Energy Physics 

NRC "Kurchatov Institute", Protvino, RF
\end{center}
\begin{abstract}
We give a new and simple proof of the inconsistency of the Bethe-West-Yennie parametrization for Coulomb-nuclear interference.
\end{abstract}

\section*{Introduction}	
65 years ago, in 1958, Bethe \cite{Be}, when analyzing the influence of the Coulomb interaction on the pion-nucleus interaction (in a non-relativistic context), derived on the basis of the eikonal representation and a number of assumptions\footnote{ For example, additivity of the eikonal with respect to the strong and electromagnetic interactions.}, proposed the following parametrization for the full scattering amplitude of charged hadrons
\begin{equation}
T_{C+N} = T_{N} + e^{i\alpha\phi}T_{C}.
\end{equation}
Here $T_{N}(s,t)$ stands for the pure strong interaction ("nuclear") scattering amplitude with usual Mandelstam variables $ s $ and $ t= -q^{2} $  while $ T_{C} $ is the Coulomb scattering amplitude  (e.g., for $ pp $ scattering)
\[T_{C} = -\frac{8\pi\alpha s}{q^{2}}F^{2}(q^{2})\] with the e.m. form factor $ F $, $ \alpha $ is the standard notation for the fine structure constant.
Since then, and especially after the modification of the phase $ \phi $ by West and Yennie \cite{We} and more detailed account for the e.m. form factor by Cahn\cite{Ca},  parametrization (1) has been unconditionally used in the vast majority of works, both experimental and theoretical, dealing with Coulomb-nuclear interference. 

The initial "Coulomb phase" derived by Bethe was essentially
\begin{equation}
\phi(s,q^{2}) = ln\frac{2}{B(s)q^{2}}.
\end{equation}

The West-Yennie modified this expression to
\begin{equation}
\phi(s,q^{2}) = ln\frac{2}{B(s)q^{2}} - \textbf{C},
\end{equation}
and finally $ \phi $ was refined by Cahn to the form
\begin{equation}
\phi(s,q^{2}) = ln\frac{2}{B(s)q^{2}} - \textbf{C} - ln(1+\frac{8}{B(s)\Lambda^{2}}) + \mathcal{O}(q^{2}/\Lambda^{2})
\end{equation}
with $ B(s) $ the elastic slope of the differential cross-section, $\textbf{C} = 0.5772...$ the Euler constant  and $ \Lambda $ the scale of the dipole form factor parametrization, $ F =(1 + q^{2}/\Lambda^{2})^{-2}$.
It should be recognized that the BWY representation (1) is quite attractive due to its compactness and ease of practical use.

Nonetheless, the Bethe-West-Yennie-type parameterization was later criticized in paper \cite{Ku} (and subsequent similar papers of the Prague Group), as well as in paper \cite{Petr}. In particular, it was pointed out that the BWY parameterization is only valid if the phase of the nuclear amplitude $ T_{N} (s,q^{2})$ , 
$ \textit{Arg} (T_{N} (s,q^{2}))$, does not depend on the momentum transfer\footnote{Later it was proved \cite{Pe2} that such an independence implies the very amplitude $ T_{N} (s,q^{2})$ to  be identically zero }.

However, this criticism turned out to be not very impressive, and the use of the BWY-type parametrization continues until very recently, e.g., in recent publications of the ALFA/ATLAS group \cite{AT} and the STAR Collaboration\cite{ST}. 

It is not improbable that the argumentation in works \cite{Ku},\cite {Petr}  might seem either unconvincing or too complicated.This prompts us to give a new and simple proof of the inconsistency of the BWY-type parametrization.

Specifically, below we will provide a proof that parameterization (1) as an equation relative to the value of $ \varphi $ has no solution.

\section*{The Proof}

In order not to clutter up our derivation with fundamentally unimportant but cumbersome details, we will consider the case when the transfers are so small that we can neglect the value of $ q^{2} $ wherever this does not lead to singularities and also we take $ F =1 $ (as was initially assumed in BWY). Incidentally, we note that typical transfers, where Coulomb-nuclear interference is already significant, are of the order of $ 10^{-3} GeV^{2}$. We will also take as the nuclear amplitude $ T_{N} $ the "popular" expression used also in \cite{AT}

\begin{equation}
T_{N}(s,q^{2}) =(\rho(s) + i)\sigma_{tot}(s) e^{-B(s)q^{2}/2}
\end{equation}
where $\rho(s)= Re T_{N}(s,0)/Im T_{N}(s,0)$ and $ \sigma_{tot}(s)$ is the $ pp $ total cross-section in $ GeV^{-2} $.
To proceed further we have to deal not with the very amplitudes $ T_{C+N} $ which suffer from an IR divergency residing in the phase but with the moduli squared $ \mid T_{C+N}\mid^{2} $ which are free from these singularities\footnote{The fact that the amplitude in Eq.(1) is certainly assumed to be finite means that the diverging part has been subtracted from the common (divergent) phase. However, such a procedure leads to a phase ambiguity, which makes it difficult to compare the results for different procedures, often not even explicitly described. The amplitude moduli squared are not only finite, but completely unambiguous and experimentally observable via differential cross-section.}.

So, on the one hand, the squaring of the modulus of the amplitude (1) gives
\begin{equation}
\mid T_{C+N}\mid^{2} = \mid T_{N}\mid^{2} + 2 \cos(\alpha\phi) ReT_{N}T_{C} + 2\sin (\alpha\phi)Im T_{N}T_{C} + T_{C}^{2}.
\end{equation}
 On the other hand, the general expression to all orders in $ \alpha $ reads \cite{Web} 
 $$
\mid T_{C+N}\mid^{2} =
\mid T_{N}\mid^{2}
\cdot
\left| ~\Gamma (1+i\alpha)~
_{1}F_{1}(i\alpha,1;z)~\right|^{2} +
~~~~~~~~~~~~~~~~~~~~~~~~~~~~~~~~~~~~~~~~~~~~~~~~~~
$$
\begin{equation}
+T_{C}\cdot \mbox{Re}\left\{ T^{*}_{N} \cdot \exp\left[ i\alpha
 \ln\left(\frac{2}{Bq^{2}}\right)\right]
\cdot\Gamma (1+i\alpha)~
_{1}F_{1}(i\alpha,1;z) \right\} +
T_{C}^{2}
\end{equation}
where $ z = Bq^{2}/2 $ and $_{1}F_{1}(i\alpha,1;z)$ is one of the confluent hypergeometric
functions\footnote{~\vspace{-4.1mm}
$$
_{1}F_{1}(i\alpha,1;z)=
\frac{1}{\Gamma(i\alpha)\Gamma (1-i\alpha)}\int_{0}^{1}
dx e^{zx} x^{i\alpha -1} (1-x)^{-i\alpha} .
$$
}
(see Chapter 9.21 in \cite{Grad}.)

Certainly, a direct extraction of $\phi $ from Eqs.(6) and (7) is 
inconceivable.
So, we follow Fermi's advise : "When in doubt, expand in a power series", do expand both Eqs(6) and (7) in series in terms of the fine structure constant $ \alpha $ and then equate the coefficients at the same powers of $ \alpha $ .
If the formula (1) is correct then extracting $ \phi $ from the first, second etc, orders we have to obtain the same value.
At small $ q^{2} $ and $T_{N}$ as in (5) the equation $ r.h.s(6) = r.h.s.(7) $ simplifies to
\begin{equation}
\cos (\alpha\phi - \Phi_{N}) = \cos(\alpha L - \Phi_{N})Re \Gamma (1+i\alpha) - \sin (\alpha L - \Phi_{N}) Im
 \Gamma (1+i\alpha)  
\end{equation}
where 
\[\Phi_{N}= \arctan \frac{1}{\rho(s)}, \]
\[L = ln\frac{2}{B(s)q^{2}}.\]
The first order gives
\begin{equation}
\phi = L - \mathbf{C}
\end{equation}
with $ L = \ln \frac{2}{Bq^{2}} $.
This exactly coincides with the West-Yennie expression (2)\cite{We}.
However, the comparison of the second order is already disturbing:
\begin{equation}
\phi^{2} = (L - \mathbf{C})^{2} - \pi^{2}/6 \approx (L - \mathbf{C})^{2} - 1.645
\end{equation}
instead of just $ (L - \mathbf{C})^{2} $.
The result of investigation of the third order looks even weirder:
\begin{equation}
\phi^{3} = (L - \mathbf{C})^{3}-L(6\zeta(3)+\pi^{2}/2) + \mathbf{C}(\pi/3 - 6\zeta(3)) - \mathbf{C}^{2}(1+\pi^{2}/6) +\mathbf{C}^{2}+\zeta(3)
\end{equation}
\[\approx(L - \mathbf{C})^{3}-12.135L -1.744\]
instead of just $ (L - \mathbf{C})^{3} $ and so on.

In other words, Eqs.(9) -(11) show that Eq.(8) (equivalent to Eq.(1)) as an equation for the quantity $ \phi $ independent of $ \alpha $ has no solution.

\section*{Conclusion}

Thus, we see that the use of the Bethe-West-Yennie formula (1) leads to a fatal ambiguity of the parameter $ \phi $ and this conclusion still holds with a non-trivial form factor either.
This means that the equation for the BWY phase has no reasonable solution at least if the nuclear amplitude $ T_{N} $ is of the form (5). 

It is worth noting that the correct general formula for taking into account the Coulomb-nuclear interference and with no ambiguities is given in papers \cite{Web} and \cite{Pst}. 
 \section*{Acknowledgements}

I am grateful to Anatolii Likhoded, Nikolai Tkachenko and the reviewer for useful and constructive discussions.

\textit{Note added in proofs}. Note that, as indicated by the reviewer, the full expression for $ T_{C+N} $ can be reduced ( by redistributing terms depending on $ \alpha $) to the expression formally similar to Eq.(1). This consists of a replacement of  the nuclear (independent on $ \alpha $) amplitude $ T_{N} $ with an amplitude $ T^{'}_{N}(\alpha) $ containing the electric charge.
In principle, such a trick could provide  an unambiguous value for $ \varphi $ and to remain in agreement with the correct results \cite{Web} and \cite{Pst}.
However, the practical value of such a replacement would be negligible compared to the BWY parametrization.

\end{document}